\documentclass[10pt,conference]{IEEEtran}

\usepackage[style=ieee]{biblatex} 

\bibliography{references.bib}    
\usepackage[normalem]{ulem}
\useunder{\uline}{\ul}{}
\usepackage{biblatex}
\usepackage[bookmarks=false]{hyperref}
\usepackage{amsmath}

\usepackage{url}
\usepackage{graphicx}  
\usepackage{float}  
\usepackage{tabu}
\usepackage{longtable}
\usepackage{ wasysym }
\usepackage{amssymb}
\usepackage{pifont}
\usepackage{tabularx}
\usepackage{url}
\usepackage{lipsum}
\usepackage{enumitem}

\usepackage[utf8]{inputenc}

\newcommand{\shorten}[1]{}
\IEEEoverridecommandlockouts
\begin{document}
\title{A Cloud-based SDN/NFV Testbed for End-to-End Network Slicing in 4G/5G}
\author{\IEEEauthorblockN{Ali Esmaeily, Katina Kralevska, and Danilo Gligoroski
}
\IEEEauthorblockA{
Dep. of Information Security and Communication Technology, Norwegian University of Science and Technology (NTNU)\\
Email: \{ali.esmaeily, katinak, danilog\}@ntnu.no
}
}
\maketitle

\begin{abstract}
Network slicing aims to shape 5G as a flexible, scalable, and demand-oriented network. Research communities deploy small-scale and cost-efficient testbeds in order to evaluate network slicing functionalities. We introduce a novel testbed, called 5GIIK, that provides implementation, management, and orchestration of network slices across all network domains and different access technologies. Our methodology identifies design criteria that are a superset of the features present in other state-of-the-art testbeds and determines appropriate open-source tools for implementing them. 5GIIK is one of the most comprehensive testbeds because it provides additional features and capabilities such as slice provision dynamicity, real-time monitoring of VMs and VNF-onboarding to different VIMs. We illustrate the potentials of the proposed testbed and present initial results.
\end{abstract}

\section{Introduction}
Network slicing is considered as an enabling technology to achieve the ambitious expectations of the fifth generation of mobile networks (5G), such as providing various services with different requirements over the same network. An end-to-end (E2E) network slice \cite{7926921} is an isolated logical network that provides a specific network service based on an accurately defined service demand upon a shared physical infrastructure. Each slice can then be controlled and managed independently. 

\begin{figure}[h]\label{NFV MANO}
\centering
\vspace{-0.2cm}
\includegraphics[scale=0.4]{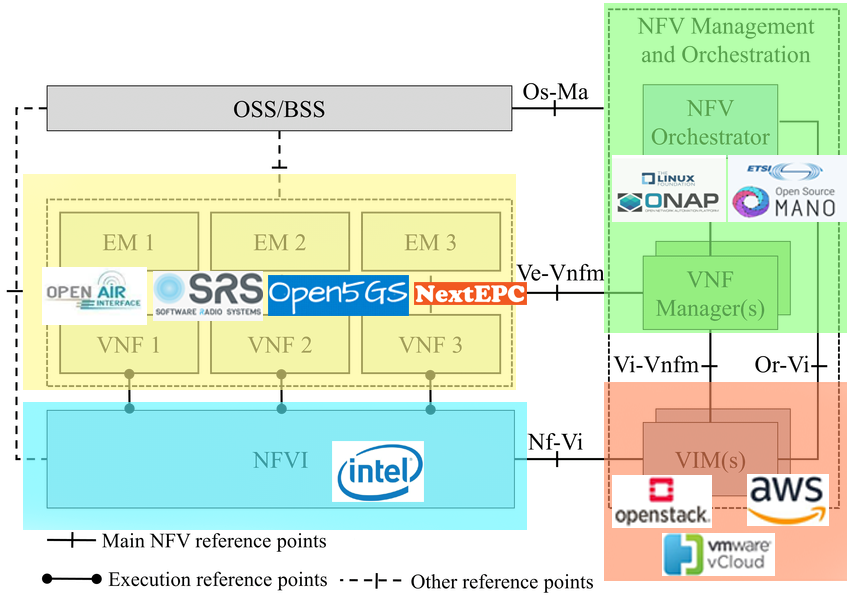}
\vspace{-0.6cm}
\caption{Different open-source software solutions mapped to the ETSI-NFV MANO framework~\cite{etsi2013network}.}
\vspace{-0.2cm}
\end{figure}

Network Function Virtualization (NFV) \cite{8320765}, Software Defined Networking (SDN) \cite{a4} and Cloud computing \cite{8320765} are the three key technologies for implementing network slicing in 4G/5G.
Since 5G aims to provide ultra-low latency services, Multi-access Edge Computing (MEC) is a complementary technology to cloud computing. 
In addition to these enabling technologies, the existence of an entity which performs efficient resource management and orchestration is inevitable. The Management and Network Orchestration (MANO) framework coordinates between available physical and virtual networking, storage and compute resources. These resources are required for creating, managing and delivering services through different slices. ETSI has developed a NFV MANO framework \cite{etsi2013network} that is composed of three functional blocks connected via reference points presented in Figure \ref{NFV MANO}:
\begin{itemize}[leftmargin=*]
\item Virtualized Infrastructure Manager (VIM) controls the NFV Infrastructure (NFVI) resources within an operator’s infrastructure domain. Thus, VIM is able to gather performance and fault measurement information of these resources. VIM also oversees the allocation of the NFVI resources to the available Virtual Network Functions (VNFs).
\item VNF Manager (VNFM) supervises a VNF or multiple VNFs and performs the life cycle management of VNF instances. Life cycle management includes setting up, maintaining and taking down VNFs.
\item NFV Orchestrator (NFVO) manages resource and service orchestration and is responsible for the entire life cycle management of various network services. Firstly, NFVO collects information about physical and virtual resources located in NFVI via the VIM. Secondly, NFVO updates its information about the available VNFs in NFVI continuously. In this way, NFVO initializes several network services by chaining particular PNFs and/or VNFs. NFVO can maintain and terminate a network service whenever there is no call for that specific service.
\end{itemize}
\vspace{-0.3cm}

One cost-efficient way of testing new solutions in 5G is by developing testbeds with functionalities close to real networks. 
By cost-efficient, we mean that setting up the testbed does not require the purchase of specialized purpose hardware and software, i.e. the software solutions are easily deployable and maintainable on standards PCs.
Figure \ref{NFV MANO} shows some of the leading open-source solutions for the different modules in the NFV MANO framework. 
A network slice can span over all network domains: 
1) Radio Access Network (RAN) emulated with Software Radio Systems LTE (srsLTE), OpenAirInterface (OAI) or Open5GS; 2) Transport Network (TN); 3) Core Network (CN) implemented with OAI or NextEPC. The MANO entity, represented with Open Source MANO (OSM) or Open Networking Automation Platform (ONAP), coordinates the available resources. OpenStack, AWS or WMware vCloud suite can act as the VIM.


\textbf{Our contribution:} The contribution of this paper is threefold. First, we combine the open-source solutions: srsLTE for the RAN, OAI for the CN, OSM for the MANO entity and OpenStack for the VIM, in order to build an E2E network slicing testbed called 5GIIK\footnote{As an online addition to this article, the source code and instructions for setting up the components are available at \url{https://github.com/aes5001/5GIIK-testbed/}
}. The value of the presented testbed is in its impact as a tool for experimentation around the community. Second, we identify the desired features for novel 4G/5G testbeds, including multi-tenancy support, orchestrating capabilities, support of multi-radio access technologies, deployment of E2E network slicing and open-source. Last but not least, we give an overview of common small-scale, non-federated and cost-efficient testbeds that are easily deployable without requiring a substantial financial investment. These testbeds are compared based on the identified features, and the comparison serves as a selection criterion for the proposed testbed.  We show that our testbed is one of the most comprehensive testbeds because it enables additional capabilities such as slice provision  dynamicity, real-time  monitoring of VMs and VNF instantiation to different VIMs. 


\section{Small-scale testbeds for network slicing} \label{stateOfArt}

\subsection{Design criteria for network slicing testbeds}
We first identify the key attributes for creating a 5G testbed that can emulate the principal features of a real network. These attributes are later used as assessment criteria for state-of-the-art testbeds.
\begin{itemize}[leftmargin=*]
\item \textbf{Support of the main enabling technologies:} The proposed testbed should be based on SDN, NFV and cloud computing. Therefore, flexibility, programmability and dynamicity in the network are granted, in addition to providing an orchestration on different network levels.
\item \textbf{Multi-domain support:} A 5G testbed should span across all network domains (air interface, RAN, TN, CN) in order to provide realization and management of E2E network slicing. 
\item \textbf{Multi-radio access technologies support:} 5G integrates different radio access technologies (RATs), thus, Long Term Evolution (LTE), WiFi and New Radio (NR) should be deployed on the same platform. 
\item \textbf{Multi-tenancy support:} 
Several mobile network operators or service providers (over-the-top players) to be able to share infrastructure by each of them acquiring one or several network slices. 
\item \textbf{End-to-end network slicing:} The slicing should be deployed across all network domains, i.e. a network slice instance consists of network slice subnet instances from different domains \cite{8698758}.
\item \textbf{Open-source:} The testbed is open-source with well-defined interfaces.
\end{itemize}

\begin{table*}[!ht]
\vspace{-0.1cm}
\caption{Comparison of small scale testbeds for network slicing in 5G.}
\vspace{-0.4cm}
\begin{center}
\begin{tabular}{ | m{4.6cm} | m{0.7cm}| m{0.7cm}| m{0.9cm} | m{0.9cm}| m{0.9cm}| m{0.9cm}| m{0.9cm} | m{0.9cm} | m{0.9cm} | m{0.9cm} | } 
  \hline
  \textbf{Testbed} & \textbf{SDN} & \textbf{NFV} & \textbf{Cloud comp.} & \textbf{Multi-domain} & \textbf{Multi-tenancy} & \textbf{MANO} & \textbf{Multi-RATs} & \textbf{E2E slicing} & \textbf{Open-source}\\ 
  \hline
  1. Secure 5G4IoT Lab \cite{8405499}& \checked & \checked & \checked & \checked & \checked & - & - & - & - \\ 
\hline
  2. 5G Test Network (5GTN) \cite{a6}& \checked & \checked & \checked & \checked &  - & \checked & \checked & - & \checked \\ 
\hline
  3. 5G Tactile Internet platform \cite{8718538}& \checked & \checked & \checked  & \checked &  \checked & \checked & - & \checked & \checked \\ 
  \hline
  4. Mosaic5G \cite{Nikaein:2018:MAF:3276799.3276803}& \checked & \checked & \checked  & \checked &  \checked & \checked & \checked & \checked & \checked \\ 
  \hline
  5. Orion \cite{Foukas:2017:ORS:3117811.3117831}&  \checked & \checked & \checked & \checked &  \checked & \checked & - & - & - \\ 
  \hline
  6. 5G Testbed for Network Slicing \cite{8656861}& - & \checked &  \checked & \checked & - & - & - & \checked & \checked\\
  \hline
  7. POSENS \cite{8524891}& \checked & \checked & \checked  & \checked &  \checked & \checked & \checked & \checked & \checked \\ 
  \hline
  8. UPC University testbed \cite{a12}& \checked & \checked & -  & \checked &  \checked & - & \checked & - & - \\ 
  \hline
  9. M-CORD based 5G Frameworks \cite{8631816}&\checked & \checked & \checked  & \checked &  \checked & \checked & \checked & \checked & \checked \\ 
  \hline
  10. NS for 5G IoT and eMBB \cite{8627115}& \checked & \checked & \checked  & \checked &  \checked & - & \checked & - & - \\ 
  \hline
  11. CHARISMA \cite{7980670}& \checked & \checked & \checked  & \checked &  \checked & \checked & - & \checked & \checked \\ 
  \hline
  12. Slice-Aware Service Assurance \cite{8806679}& \checked & \checked & - & - & - & \checked & - & - & - \\ 
  \hline
  13. Simula Metropolitan Centre \cite{10.1007/978-3-030-44038-1_105}& \checked & \checked & \checked & \checked & \checked & \checked & \checked & - & \checked \\ 
  \hline
  \textbf{14. 5GIIK (our proposal)} &\checked & \checked & \checked  & \checked &  \checked & \checked & \checked & \checked & \checked \\ 
  \hline
\end{tabular}
\end{center}\label{TableOverview}
\vspace{-0.6cm}
\end{table*}

\subsection{An overview of the state-of-the-art network slicing testbeds}
We give here a short description of the existing testbeds and compare them in Table \ref{TableOverview}.

\shorten{
\subsubsection{OpenAir5G Lab \cite{Nikaein:2014:OFP:2677046.2677053}}
}

\subsubsection{Secure 5G4IoT Lab \cite{8405499}} This testbed deploys OAI in containers to virtualize both Evolved Packet Core (EPC) and eNB. For scalability purposes, the testbed has been implemented in several containers. 
It has been evaluated by producing two isolated network slices (eHealth and Internet light) on the same infrastructure. 

\shorten{
}

\subsubsection{5G Test Network (5GTN) \cite{a6}} In this testbed, located at Oulu University, the RAN operates on licensed LTE and 5G bands. 
By changing the Access Point Name between EPC (deployed on OpenStack) and IP Multimedia System (deployed on VMWare), UE switching between two slices is possible. The testbed has been tested for CPU utilization, throughput and delay for the two specific slices.

\subsubsection{5G Tactile Internet platform \cite{8718538}} This testbed follows the SEMIoTICS architecture, consisting of backend/cloud, networking and field layers, to create a 5G platform for providing E2E services for industrial IoT applications with sub-millisecond latency. 
The testbed performance has been assessed for performing E2E slicing and dynamically sharing the available bandwidth between two VNFs, one for smart monitoring and one for actuating.

\subsubsection{Mosaic5G \cite{Nikaein:2018:MAF:3276799.3276803}} This testbed brings flexibility and scalability to service provision. The testbed architecture consists of five software modules along with hardware components: OAI, FlexRAN, LL-MEC, Store and JOX. 
The Mosaic5G platform has been used for a few use cases such as critical e-Health, V2X communication for intelligent transportation systems and multi-service management/orchestration for smart cities.

\subsubsection{Orion \cite{Foukas:2017:ORS:3117811.3117831}} 
The architecture of Orion provides the sharing of RAN resources in addition to providing isolation between slices, and so, operation in one slice cannot degrade the performance of another slice. This is achieved by having an independent control plane in the RAN domain for each slice. As a result, Orion offers the opportunity to deploy different service characteristics in the RAN domain and it is a concrete step towards realizing RAN-as-a-Service. 

\subsubsection{5G Testbed for Network Slicing Evaluation \cite{8656861}} The testbed utilizes OAI for both RAN and CN domains. There are two CNs which share radio resources of a single eNB in the RAN. 
The testbed has been appraised for connection establishment for both normal LTE UEs and UEs with an implemented Network Slice Selection Assistance Information. 

\subsubsection{POSENS \cite{8524891}} 
POSENS provides efficient resource utilization for creating independent and customizable E2E slices. RAN slicing can be realized via three possibilities: 1) Slice-aware shared RAN where the whole radio domain is shared, but CNs are distinguished by the specific services they provide and a UE can utilize different slices provided by the CNs; 2) Slice-specific radio bearer where only cell-specific functionality is shared; and 3) Slice-specific RAN where apart from the air interface, slices of different tenants are isolated in other protocol stack layers. 

\subsubsection{UPC University testbed \cite{a12}} This testbed implements automatically RAN slicing via RESTful API. The testbed applies the slice-aware policy in Radio Resource Management (RRM) for admission control and scheduling processes. 5G-EmPOWER \cite{a18}, as the central entity in the testbed, allows RAN slicing management and it also shares the available radio resources among the created RAN slices according to RRM descriptors.

\subsubsection{Mobile-Central Office Re-Architected as Datacenter (M-CORD) based 5G framework \cite{8631816}} The work in \cite{8631816} focuses  mainly on OAI integration with the M-CORD platform and different implementation procedures to deploy LTE network on top of M-CORD. 

\subsubsection{Dynamic Network Slicing for 5G IoT and eMBB services \cite{8627115}} This testbed demonstrates the sharing of the same RAN resources among enhanced Mobile Broadband (eMBB) and IoT services. 
The real-time slicing decision in C-RAN is performed by a SDN controller (FlexRAN) that connects via its Northbound Interface to an entity called Slicing app, which includes IoT and eMBB modules. 

\subsubsection{CHARISMA testbed \cite{7980670}} This testbed has been designed for practical analysis in the CHARISMA project, and the goal is to bring the network processing close to the users. 
By employing Ethernet Virtual Connections and Virtual LAN-ID concepts in the testbed, multi-tenancy and slice isolation are achieved. WiFi technology and cloud-based servers present the RAN and CN domains, respectively.

\subsubsection{Slice-Aware Service Assurance Framework \cite{8806679}} This testbed measures Quality of Experience (QoE) of a specific service according to the several service dependability Key Quality Indicators (KQIs). The testbed provides web content browsing and adaptive video streaming services to assess infrastructure performance and the KQIs alteration for each service.

\subsubsection{Simula Metropolitan testbed \cite{10.1007/978-3-030-44038-1_105}} This testbed demonstrates the deployment of OAI-EPC as a VNF on a cloud environment, and it presents the LTE CN service instantiation via OSM. 
The goals of this implementation are to produce MEC services to EPC and to integrate EPC with the extended eNB software. The functionality of the testbed is evaluated for establishing TCP and SCTP connections for downloading from server to UE, uploading from UE to a server, and bidirectional communication between UE and server.

As a summary of Table \ref{TableOverview}, we can classify the presented testbeds into two main categories. The first category includes those testbeds which provide some capabilities in network slicing; however, they do not attain to all of the design criteria for realization of E2E network slicing. While \cite{8405499, a6, Foukas:2017:ORS:3117811.3117831, a12, 8627115, 8806679, 10.1007/978-3-030-44038-1_105} focus on one specific network domain and provide slicing just for that particular domain, \cite{8656861} provides E2E slicing but does not have a separate entity for management and orchestration. Papers \cite{8718538, 7980670} offer only a light MANO implementation. The solutions in \cite{8718538, 8656861, 7980670} deploy E2E network slicing with MANO capability but without multi-RATs or multi-tenancy support. 
The testbeds in \cite{Nikaein:2018:MAF:3276799.3276803, 8524891, 8631816} belong to the second category, which complies with all of the designing principles. Nevertheless, 5GIIK offers more features and capabilities such as slice provision dynamicity, real-time monitoring of VMs and VNF-onboarding to different VIMs, which differentiate it from other testbeds in the second category. Section \ref{Applicability} describes the 5GIIK features.

\section{5GIIK - Our Proposed Testbed}
\label{testbedArchitecture}
We consider all features elaborated in the previous section and propose 5GIIK - a testbed architecture that grants an E2E network slicing with MANO capability, that  supports multi-tenancy and multi-RATs and at the same time it is a cost-efficient design. In this section, we first present the network architecture with all associated components, and then we explain the network slice creation and instantiation in 5GIIK.
\subsection{5GIIK testbed architecture}
\label{5GIIK testbed architecture}
\begin{figure}[t]
\centering
\includegraphics[scale=0.4]{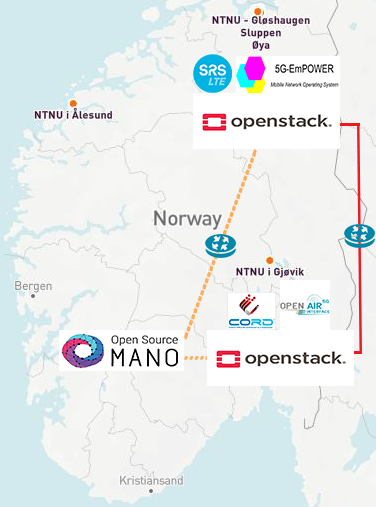}
\vspace{-0.2cm}
\caption{The used open-source solutions mapped to testbed premises in Trondheim and Gjøvik.}\label{5GIIK-Testbed-Figure-Geographic}
\vspace{-0.5cm}
\end{figure}

\begin{figure*}
\centering
\includegraphics[scale=0.40]{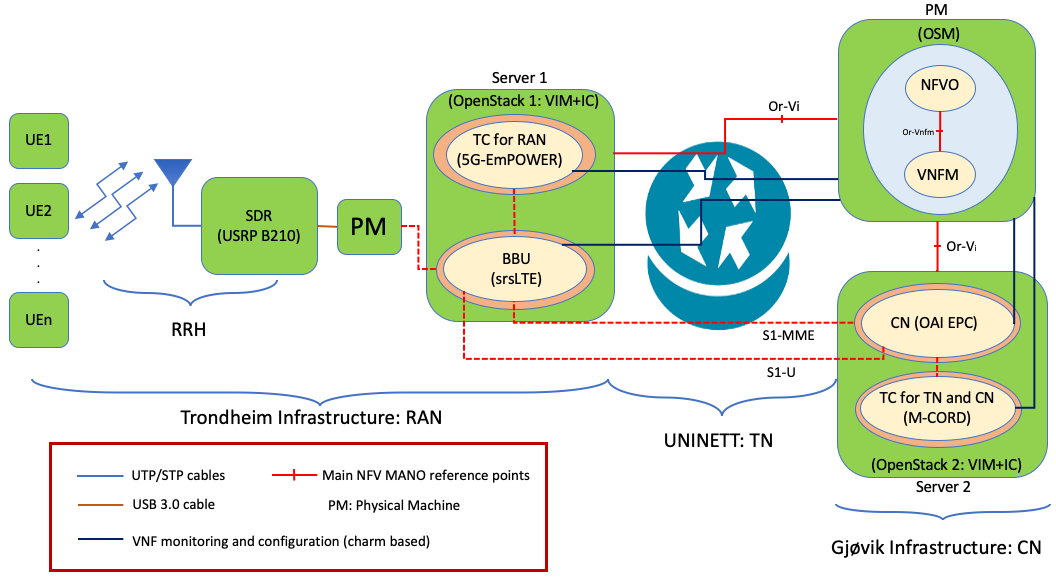}
\vspace{-0.3cm}
\caption{5GIIK testbed architecture.}
\vspace{-0.1cm}\label{5GIIKTestbedArchitecture}
\end{figure*}

A high-level description of the 5GIIK testbed is given in Figure \ref{5GIIK-Testbed-Figure-Geographic}.
It is composed of several entities that emulate real 4G/5G networks. The RAN and CN parts of the testbed are implemented in Trondheim and Gjøvik campuses of NTNU, respectively. The IP backbone network, which is provided by Norway's National Research and Education Network (UNINETT), is used as TN in our platform. Our testbed virtualizes not only the CN but also the Base Band Unit (BBU) of RAN into the cloud to build a Cloud-RAN (C-RAN) architecture. 
The remote radio head section of the C-RAN is managed by a Software Defined Radio (SDR) and connected antennas to the SDR. 
The architecture, illustrated in Figure \ref{5GIIKTestbedArchitecture}, is partially motivated by the work in \cite{7926921}, \cite{8405499} and \cite{8631816}. To simplify, the connections of NFV MANO reference points with external entities are not depicted. Next, we evaluate the different open-source solutions for modules in the NFV MANO framework in Figure \ref{NFV MANO} and explain our reasoning behind the selected solution for the proposed architecture in Figure \ref{5GIIKTestbedArchitecture}.
\begin{itemize}[leftmargin=*]
\item \textit{VNFs:} OAI is a flexible solution for emulating LTE systems and implements the full protocol stack of 3GPP standard in Evolved-UMTS Terrestrial Radio Access Network (E-UTRAN) and EPC. OAI can be used to build a complete LTE network (eNB, EPC and UE) on a PC or Virtual Machine (VM). NextEPC\footnote{\url{https://nextepc.org/},\ \  \url{https://open5gs.org/},\ \  \url{https://www.onap.org/}, \\ \url{https://www.vmware.com}, \url{https://aws.amazon.com/}, \url{https://opencord.org/}, \url{https://www.openstack.org/}, \url{https://www.softwareradiosystems.com/},\\ \url{https://osm.etsi.org/}, \url{https://www.opennetworking.org/onos/}}, as its name indicates, implements just the CN of a 4G/5G system. Open5GS$^1$ provides a complete implementation of a 4G/5G. However, the lack of detailed documentation about its specifications is a repelling point for choosing it. The srsLTE solution emulates the whole system and has a well-structured code for future improvements. The srsLTE library is modular and utilizes single instruction multiple data operations for increasing its performance in the system. It has a light implementation regarding the CN. From a hardware perspective, the srsLTE library can operate with various front end RFs and it is able to provide interfaces for different types of Ettus USRP pieces of equipment. 
\item \textit{NFV orchestrator:} There are various solutions, but the main competition is between ONAP$^1$ and OSM. Considering the compatibility of these orchestrators with different VIMs, OSM supports several VIMs and can manage them at the same time. Regarding resource usage (CPU and memory), again, OSM bests ONAP by utilizing fewer resources compared to ONAP \cite{DBLP:journals/corr/abs-1904-10697}.
\item \textit{VIM:} There are solutions such as OpenStack, VMware vCloud Director$^1$ and Amazon Web Services$^1$. OpenStack is more potent than others for infrastructure orchestration, scalability and resource utilization. Besides, OpenStack can be deployed on standard machines with the appropriate amount of resources.
\end{itemize}

Following this elaboration, 5GIIK testbed uses OAI as CN and srsLTE as RAN. The MANO entity in 5GIIK is represented by OSM, which is developed in Python and operates on Linux. OSM combines NFVO, VNFM of the NFV MANO architecture. As a result, configuration and abstraction of VNFs, orchestration, and the management of the network services are feasible. Since release 4, OSM uses Docker container technology and cloud-based solutions. 5GIIK also integrates SDN controllers to its architecture. 
SDN-based Tenant Controller (TC) is needed to provide L2 VLANs to manage tenant VNFs located in different VIMs while implementing network slicing in TN. We integrate two TCs for the whole network domains to make our design more generic. 

\subsubsection{5G-EmPOWER as TC for RAN domain}
5G-EmPOWER controller \cite{a18}, also known as EmPOWER, is an open multi-access network operating system. It is created upon a single platform that consists of general-purpose hardware (x86) and Linux. 5G-EmPOWER is RAT-agnostic and by using a data plane programmability policy, it manages the virtualized network resources of multiple radio nodes (WiFi Access Points, LTE eNBs and 5G NR). It communicates with the eNB in the C-RAN to manage and control the radio resource allocation to the end-users. Consequently, it supports multiple virtual networks (tenants) on top of the same physical infrastructure. The latest version of 5G-EmPOWER is compatible with srsLTE eNB, and it fits well in our testbed.
\subsubsection{M-CORD as TC for TN and CN domains} M-CORD$^1$ is a cloud-based solution built on SDN, NFV technologies. It encompasses both virtualization of RAN functions (vRAN) and a virtualized CN (vEPC) to allow mobile edge applications and services using a micro-service architecture. M-CORD disaggregates and virtualizes network functions and operator services. By integrating the ONOS$^1$ controller in its architecture, M-CORD implements network slicing in TN in order to form an E2E network slice.

For deploying E2E network service orchestration, OSM interacts with VIMs via Or-Vi interfaces. OSM performs lifecycle management of NF configuration, operation and monitoring by interacting with Physical/Virtual NFs (EPC, BBU) via charm configuration files. The infrastructure management section of NFV MANO is divided into VIM and Infrastructure SDN based Controller (IC), represented in both OpenStack 1 and 2 in Figure \ref{5GIIKTestbedArchitecture}. VIM, in cooperation with IC, manages and controls the infrastructure layer, both physical and virtual resources, via Nf-Vi. 

\subsection{Network slice instantiation in 5GIIK testbed}
\label{Slice provisioning}
\begin{figure*}[h!]
\centering
\vspace{-0.2cm}
\includegraphics[scale=0.6]{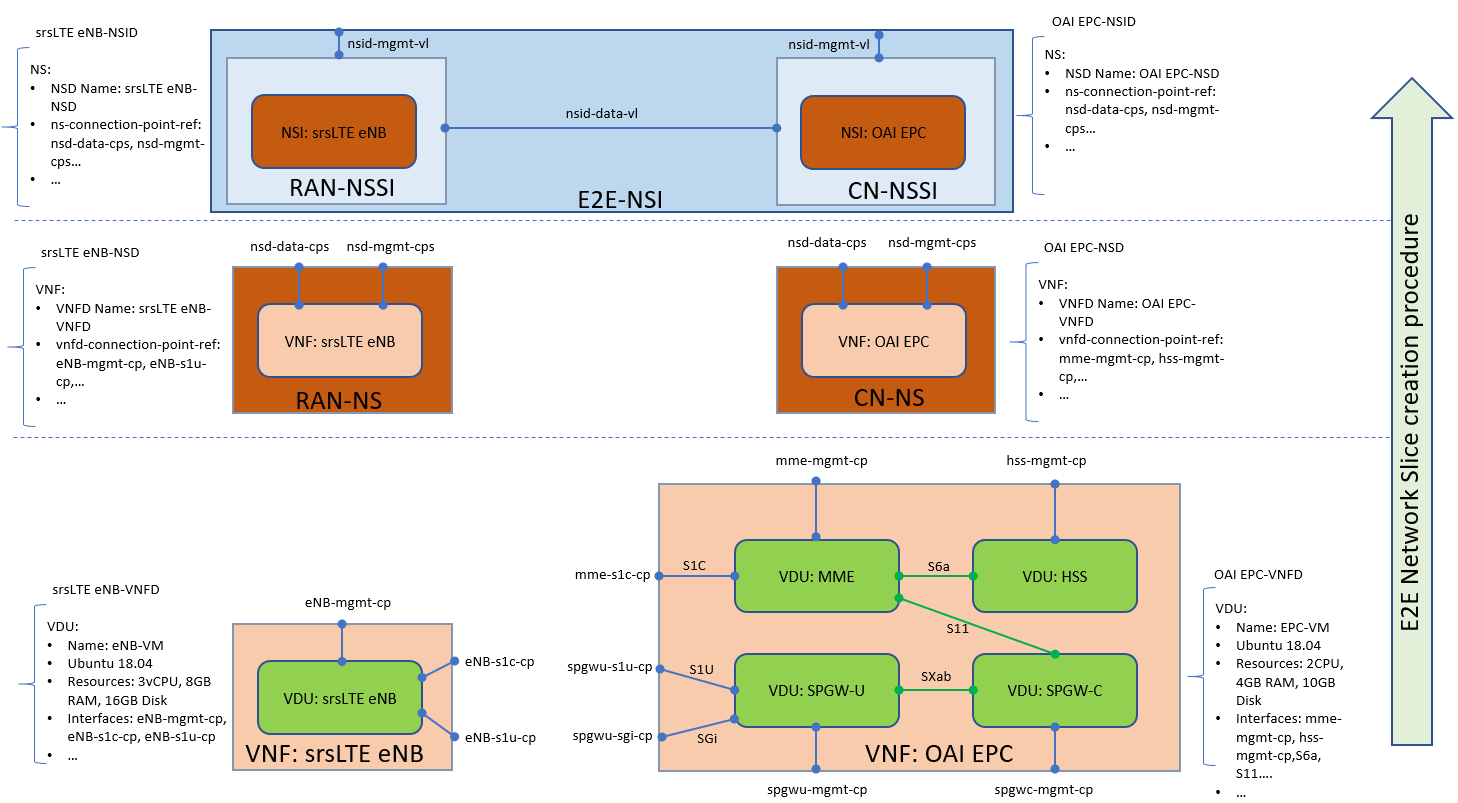}
\vspace{-0.3cm}
\caption{Procedure of E2E network slice creation.}
\vspace{-0.1cm}
\label{NetSliceCreation}
\end{figure*}

Now we describe how a network slice is instantiated by the OSM in the 5GIIK testbed, and the procedure is illustrated button-up in Figure \ref{NetSliceCreation}. 

VNF Descriptors (VNFDs) are located in the first level of creating network slices. VNFD is a file that retains the information such as the software image that the VNF needs to be built on as well as CPU, memory and storage that the VNF needs for high-performance operation, internal virtual links (vls) between Virtualization Deployment Units (VDUs) inside a VNF and a lifecycle event of a network slice. A management network (mgmt) is needed to assign IP addresses to the launched VDUs. In this level, two VNFDs are required; OAI EPC-VNFD and srsLTE eNB-VNFD. 
OAI EPC contains four entities: Home Subscriber Server (HSS), Mobility Management Entity (MME), Control Plane of the Service Packet Gateway (SPGW-C) and User Plane of the Service Packet Gateway (SPGW-U). Thus, OAI EPC-VNF includes four VDUs, one for each entity, while srsLTE eNB-VNF includes one VDU in the context of OSM.

Network Service Descriptors (NSDs) are positioned in the second level of the slice formation. NSD comprises different launched VDUs that a network service needs to operate. NSD also includes external connection points (cps) between the demanded VDUs. The same as the VNFD level, two NSDs are essential here; OAI-EPC-NSD and srsLTE eNB-NSD.

Finally, in the third level of a network slice creation, a Network Slice Instance Descriptor (NSID) chains the established service instances and forms a network slice. In this level, one NSID is needed to chain the launched services in the CN and RAN domains. 
\vspace{-0.1cm}
To summarize, these are the steps to instantiate an E2E network slice: 
\begin{enumerate}
    \item At the VNFD level, OSM instantiates the required VDUs (VMs) via its resource orchestrator block to the VIM. In this level, the necessary resources are allocated, and specific interfaces are configured on each VDU.
    \item  At the NSD level, according to the launched VMs, particular service instance(s) is (are) created.
    \item At the NSID, the service instances are chained and create an E2E network slice.
\end{enumerate} 

Apart from the crucial information that has to be defined on each descriptor level, some customized information can be set as well. For instance, it is possible to define some metric parameters for running VNFs to be collected via VIM. In this way, performing periodic network monitoring at the infrastructure level is achieved.

\section{5GIIK Features and Initial Experiments}
\label{Applicability}
The proposed testbed provides a whole range of features that can be exploited for developing various new solutions for network slicing in wireless and mobile networks. 
\begin{itemize}[leftmargin=*]
\item 5GIIK is cross-domain and it spans over the whole network in order to support E2E slicing.
\item The SDN functionality in 5GIIK enables studies of numerous use cases and scenarios for new resource allocation techniques. In particular, 5G-EmPOWER supports common machine learning toolkits, which is a missing capability for most of the testbeds mentioned in Table \ref{TableOverview}. Hence, slice-aware traffic marking strategies can provide dynamicity in slice provision for different use cases, and it can assign the available radio resources to the end-users in an optimized fashion.
\item 5GIIK allows multi-RAT implementation. 
The recent containerized-based 5G-EmPOWER release (under an APACHE 2.0 License) is compatible with WiFi access points and srsLTE (release 19.09) to perform RAN slicing for both of these RATs. As a result, RAN slicing for both LTE and WLAN is applicable.
\item Multi-tenancy that is fundamental in the 5G era, especially in the RAN domain, is supported in our architecture. In particular, 5G-EmPOWER grants 5GIIK to create two different tenants via its web interface. By defining a Mobile Network Operator (MNO) such as PLMN-ID=A, radio resources can be shared among two Mobile Virtual Network Operators (MVNOs) so-called foo and bar. Each of the foo and the bar has its own unique MVNO-identifier. Then, on each of these MVNOs, one or several network slices can be created. Consequently, one user equipment can be configured in such a way to connect to the desired slice on a particular MVNO. Figure \ref{multiTenancy} demonstrates two scenarios in which there are one and two MVNOs with their slices, respectively. 
 
\begin{figure}[h] 
\vspace{-0.2cm}
\centering
\includegraphics[scale=0.2]{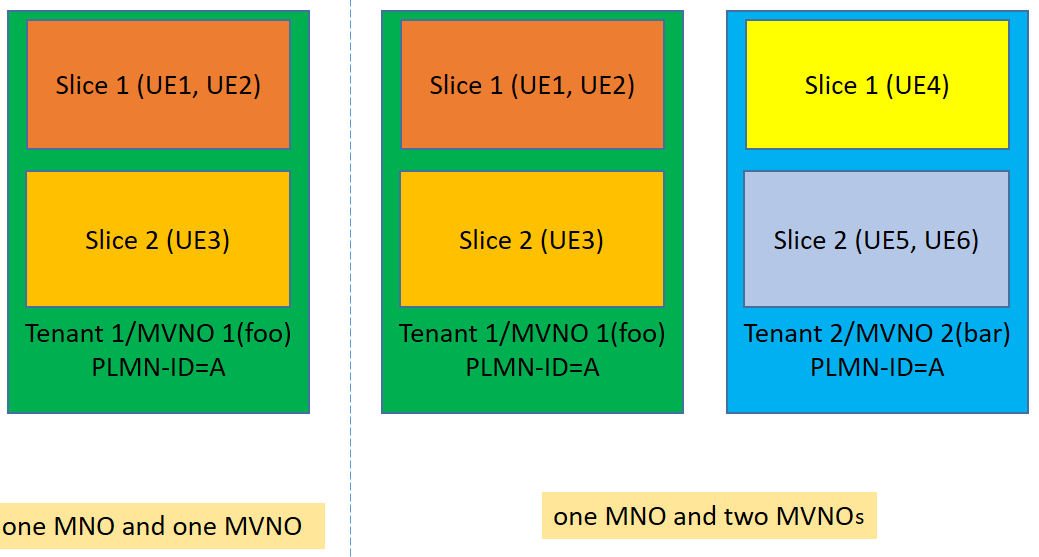}
\vspace{-0.4cm}
\caption{Multi-tenancy support in 5GIIK.}\label{multiTenancy}
\vspace{-0.2cm}
\end{figure}
\item OSM and OpenStack perform management and orchestration in 5GIIK. Besides, OSM can manage multiple-VIMs, since our testbed is implemented on two OpenStack infrastructures (RAN in Trondheim and CN in Gjøvik).
\item OSM module in 5GIIK offers E2E network slice provisioning. Firstly, VNFDs specify the desired images with their demanded resources (CPU, memory, and storage) via OpenStack. Secondly, based on the NSDs, specific service instances are created. Finally, the NSID determines how to chain these service instances to create an E2E network slice that traverses the whole network domains.
\item OSM provides the possibility to perform the VNF-onboarding process. In VNF-onboarding, the VNF lifecycle has three phases -- so-called days. In day-0, management policies for VNFs' instantiation are established. In day-1, VNFs are configured, and they can provide the demanded services. In day-2, the possibilities of reconfiguring VNFs and monitoring their Key Performance Indicators (KPIs) in runtime operation are granted. Hence, the onboarding process can be done for a variety of VNFs to build favorable VNF packages on OpenStack.
\item Prometheus is a system monitoring toolkit that can be integrated with OSM in our testbed. Specific metrics such as CPU utilization and average memory utilization can be defined at VNF descriptors. Consequently, Prometheus retrieves the collected metrics and performs real-time monitoring of all active/detective VMs.
\end{itemize}

\begin{figure*}[t]
\centering
\includegraphics[scale=0.65]{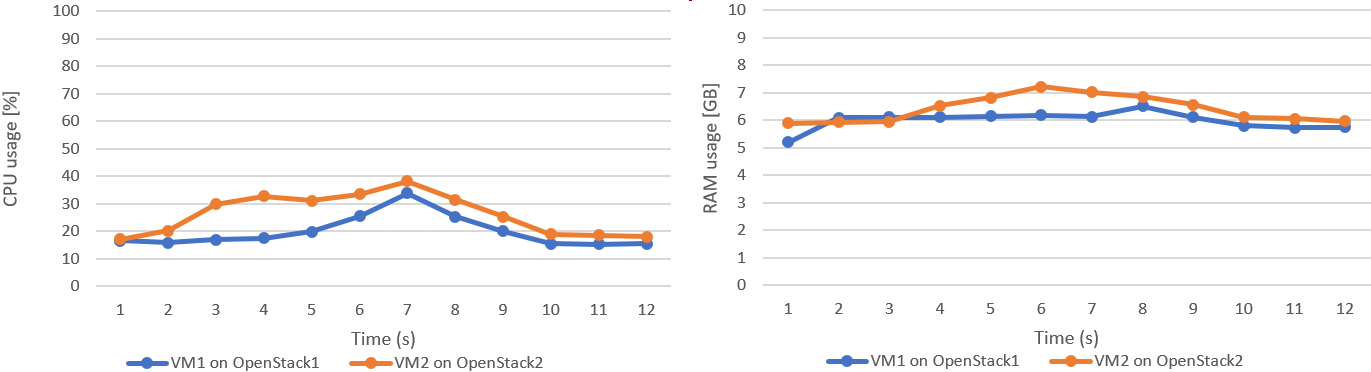}
\vspace{-0.2cm}
\caption{CPU and RAM usage during a transfer of two files from one VM to the other in 5GIIK.}\label{CPURAMBigFilesTransfer}
\vspace{-0.5cm}
\end{figure*}

\textit{Implementation challenges:} Deploying C-RAN is a challenge since the communication between BBU and RRH demands very low latency, and it is essential to implement BBU close to RRH. It is even more challenging when the network delivers delay-sensitive services with ultra-low latency requirements. Resource management in VIM(s) is considered as another challenge, mainly when a VIM is not capable of launching instances (VMs that are running VNFs) with a high amount of assigned resources in terms of CPU and memory. It is crucial to know how to allocate available resources to multiple instances in a VIM.


\subsection{Initial testing}
5GIIK is installed on three similar Intel machines (i7-4790 CPU @ 3.60GHz, 32GB RAM), which are running two OpenStack platforms and OSM. In order to evaluate the testbed performance regarding CPU and memory usage, we carried out one initial test. The test involves transferring and downloading files with different sizes (500 MB and 1 GB) from one VM in one OpenStack to another VM in the second OpenStack. First, one VNF descriptor is created on the OSM to define the required image with its demanded resources (CPU, memory and storage). Furthermore, the management network has to be set to assign an IP address to the service instance. Subsequently, one service descriptor is created to determine how the base image can launch the service instance. Considering the VNF and service descriptors, two similar service instances (two VMs running Ubuntu 16.04 with one virtual CPU, 16 GB of RAM and 20 GB of storage) are launched on the two VIMs. Figure \ref{CPURAMBigFilesTransfer} illustrates the CPU and RAM measurement. VM1 downloads the smaller file and then starts sending it to the VM2. A similar approach for VM2 takes place but with the larger file. As expected, VM1 utilizes less amount of resources (up to 33.9\% of CPU and 6,5 GB of RAM) compared to the VM2 (up to 36.2\% of CPU and 7,2 GB of RAM).

\vspace{-0.1cm}
\section{Conclusion}
\label{Conclusion}
Following the modern development of tools and technologies that enable network slicing, we composed a list of several design criteria for constructing testbeds. Then we summarized some small-scale testbeds with their main features analyzed through the criteria list that we composed. We also proposed 5GIIK - a testbed that performs E2E network slicing with the capability of management and orchestration of network resources. 5GIIK is an open-source-based architecture and its flexibility provides the opportunity to create innovative algorithms, patterns and solutions in the network slicing realm.


%


\printbibliography

\end{document}